\documentclass[aps,twocolumn,groupedaddress,showpacs]{revtex4}
\usepackage{graphicx}
\usepackage{dcolumn}
\usepackage{bm}
\usepackage{epstopdf}
\usepackage[dvipsnames,usenames]{color}
\usepackage{color}
\usepackage{float}
\usepackage{amsmath}
\usepackage{amssymb}
\usepackage{amsthm}
\usepackage{mathrsfs}

\usepackage[T1]{fontenc}
\usepackage[skip=1ex]{caption}
\usepackage{booktabs,tabularx}
\usepackage{siunitx}

\renewcommand{\thefootnote}{*}
\emergencystretch=\maxdimen
\begin{document}
\title{Second-order correlation function supported optical sensing for particle detection}

\author{T.~Wang$^{1,2,\footnote{wangtao@hdu.edu.cn}}$, C.~Jiang$^{1,2}$, J.~Zou$^{3}$, H.~Zhou$^{1,2}$, X.~Lin$^{4}$, H.~Chen$^{4}$, G.~P.~Puccioni$^{5}$, G.~Wang$^{1,2}$, and G.~L.~Lippi$^{6,\footnote{Gian-Luca.Lippi@inphyni.cnrs.fr}}$}

\affiliation{$^1$Engineering Research Center of Smart Microsensors and Microsystems of MOE, Hangzhou Dianzi University, Hangzhou, 310018, China}
\affiliation{$^2$School of Electronics and Information, Hangzhou Dianzi University, Hangzhou, 310018, China}
\affiliation{$^3$School of Communication Engineering, Hangzhou Dianzi University, Hangzhou, 310018, China}
\affiliation{$^4$Interdisciplinary Center for Quantum Information, College of Information Science and Electronic Engineering, Zhejiang University, Hangzhou 310027, China}
\affiliation{$^5$Istituto dei Sistemi Complessi, CNR, Sesto Fiorentino, Italy}
\affiliation{$^6$Universit\'e C\^ote d'Azur, Institut de Physique de Nice (INPHYNI), Nice, France}

\date{\today}

\begin{abstract}
We propose a new sensing method based on the measurement of the second-order autocorrelation of the output of micro- and nanolasers with intensity feedback. The sensing function is implemented through the feedback-induced threshold shift, whose photon statistics is controlled by the feedback level in a characteristic way for different laser sizes. The specific response offers performances which can be adapted to different kinds of sensors. We propose the implementation of two schemes capable of providing a quantitative sensing signal and covering a broad range of feedback levels: one is utilizing the evolution of g$^{(2)}$(0), the other one is the ratio between central and side peaks in g$^{(2)}(\tau)$. Laser-threshold-based sensing could, thanks to its potential sensitivity, gain relevance in biomolecular diagnostics and security monitoring.
\end{abstract}


\maketitle 

\section{Introduction and objectives}
The statistics of the radiation field, or photon statistics, begun with the seminal paper by Hanbury-Brown and Twiss for stellar size measurements~\cite{Brown1956,Hanbury-Brown1956}, has evolved to become one of the pillars of physics~\cite{Glauber1963L,Glauber1963,Mandel1965,Paul1982}. Its applications run from the detection of squeezed states and antibunching~\cite{Leuchs1986}, ghost imaging~\cite{Shapiro2012,Tao2021}, improving the sensitivity of gravitational wave detectors~\cite{Lawson2020}, particle sizing~\cite{Schulz-Dubois1987}, dynamic light scattering~\cite{Schaetzel1991}, fluid mechanics and velocimetry~\cite{Schulz-Dubois1983,Degiorgio1977}, quantum information~\cite{Mirhosseini2020,Kurochin2020}, and advanced measurement techniques~\cite{Silberhorn2007,Morais2020}, to name only a few.  

Its ability to detect extremely low light levels, down to the single photon, has rendered it the preferred technique for the characterization of the emission of nanolasers~\cite{Wiersig2009}, devices whose energetic efficiency, coupled to extremely reduced footprint and high integrability, promise a broad range of applications~\cite{Hill2014,Ma2019}, for instance in optical chips~\cite{Smit2012} or interconnects~\cite{Miller2017}. Low power consumption is, of course, very attractive for many other applications, including sensing.

A new operation regime, which precedes the threshold for continuous laser operation, has been recently identified:  emission of uncorrelated photon bursts of short duration (typically a nanosecond or less).  It has been directly or indirectly reported both in nanodevices~\cite{Lebreton2013L,Lebreton2013,Pan2016,Ota2017} as well as in microcavities~\cite{Wang2015,Wang2020a}.  The origin of these pulses is attributed to the phase space properties of the below-threshold operation where fluctuations in the energy reservoir can grow sufficiently large as to cause a temporary burst of photons before returning the device below threshold~\cite{Wang2020JMO,Lippi2021}.  
The interest for this kind of emission comes from the very low power consumption, as the laser is biased just {\it below threshold} and from the low coherence properties which stem from operation in the still incoherent region.   

This regime of operation is characterized by a peak photon number which can exceed by two orders of magnitude the photon average value, for the corresponding pump, which comes from the strong photon bunching.  The compressed emission of large number of photons is identified by a superthermal photon statistics~\cite{Wang2020a,Wang2020JMO} in the single-mode regime -- multimode superthermal emission~\cite{Redlich2016,Marconi2018} is due to mode competition and appears only on the weak mode:  the strong one and the total output show standard, subthermal photon statistics. The independence of the emitted pulses offers the advantage of avoiding interference and provides a new operation regime for semiconductor lasers, where feedback does not induce the usual complex dynamics, whose scenario is too complex for most practical applications, even in (the relatively more stable) Vertical Cavity Surface Emitting Lasers (VCSELs)~\cite{Panajotov2012}.

In the photon burst regime of operation, the laser is going to emit random light pulses. If a fraction of the pulse is scattered back into the laser, the entering photons are going to act as a delayed, self-injected incoherent signal which is going to facilitate the emission of a new pulse~\cite{Wang2019,Wang2020}. Hence, one may statistically expect the likely appearance of a new pulse in correspondence with the returning (fraction of the) pulse. This stimulation acts as a trigger which increases the pulse frequency while reducing, statistically, their amplitude. As a consequence, the degree of bunching is reduced and is detected by a reduction in the value of the autocorrelation. In other words, we can detect the presence (and characteristics) of a scatterer passing through the laser beam through a reduction in the amount of photon bunching. The use of the autocorrelation allows for a sensitive technique which can even be used at extremely low photon flux levels, thanks to the potential offered by single-photon detectors.

Although in this paper we concentrate on Class B lasers~\cite{Tredicce1985}, i.e., devices where the dynamical behaviour is described by two variables (carrier density and photon number), a strong photon bunching before threshold has been numerically seen~\cite{Roy2009,Roy2010} and then carefully investigated also in Class A devices (typically described only by the dynamics of the photon number)~\cite{Vallet2019}.  The numerical predictions are consistent with theoretical analyses~\cite{Takemura2019,Takemura2021}.  At the present time, much less is known on the equivalent features of Class A devices, given the intrinsic stiffness of their numerical simulations.  Future progress in their description may reveal interesting aspects for sensing applications.

This manuscript applies this mechanism to theoretically propose the realization of a sensor based on the measurement of the degree of superthermal bunching. After introducing the numerical tools for the computations (Section~\ref{Theo}), we discuss the threshold dynamics of micro- and nanolaser as well as the autocorrelation-based sensing (Section III). Two experimental implementations are examined in Section IV-A, and a discussion on the sensing features is offered in parts B-D of Section IV. Finally, Section V concludes the paper's discussion. 

\section{Theory}\label{Theo}
\subsection{Stochastic simulator model}
The model used in this work is a set of recurrence relations between carriers and photons, adapted to include feedback from the Stochastic Simulator (SS) scheme~\cite{Puccioni2015}. The SS considers the intracavity processes as a sequence of physical events (absorption, emission, etc.) occurring at discrete times~\cite{Puccioni2015}. This contrasts the traditional description of noise, often introduced in laser modelling through Langevin terms -- based on Gaussian statistics~\cite{Druten2000, Hofmann2000} -- which assumes the application of infinitesimally small perturbations in large numbers. This approach fails when considering very small devices, i.e., high $\beta$ lasers (where $\beta$ represents the fraction of spontaneous emission coupled into the lasing mode~\cite{Bjork1991}), given the very small number of photons present in the cavity~\cite{Elvira2011}. In these systems, the discreteness~\cite{Lebreton2013} of the variations (creation or destruction of a photon, as an integer unit) and the relatively long times which the process involves (compared to the intrinsic time scales) invalidate the conditions for the Langevin approximation. The model, discussed in detail in~\cite{Wang2020}, can be summarized as follows:

\begin{eqnarray}
\label{defN}
N_{q+1} & = & N_q + N_P - N_d - E_S \, , \\
S_{q+1} & = & S_q + E_S - L_S + S_{sp} + S_{inj,q-d}\, , \\
R_{L,q+1} & = & R_{L,q} + D_L - L_L - S_{sp} + R_{inj,q-d} \, , \\
\label{defRo}
R_{o,q+1} & = & R_{o,q} +(N_d - D_L) - L_o \, ,
\end{eqnarray}

\noindent where $N_P$ represents the pumping process, $N_d$ is the process of spontaneous relaxation, which reduces the population inversion $N$ and is split into the fraction of spontaneous photons into the lasing mode ($R_L$) and in the remainder of the cavity modes ($R_o$) through the probabilistic operator $D_L$. $E_S$ represents the stimulated emission process which also depletes the population, $L_S$ and $L_L$ represent the leakage of stimulated and spontaneous photons in the lasing mode, respectively, through the output coupler, while $L_o$ stands for the escape process for the other spontaneous photons outside the cavity volume. $S_{sp}$ provides the starting seed~\cite{Puccioni2015} for the first stimulated emission process. $S_{inj, q-d}$ and $R_{inj, q-d}$ represent the fraction of stimulated and spontaneous photons, respectively, corresponding to the delayed index $(q-d)$ where $d$ matches the delay time $2 \tau_{ext}$. Details, including the parameter values not specified here, can be found~\cite{Puccioni2015}. 

All random processes are Poissonian (implementable also with a Binomial distribution~\cite{Puccioni2015} when the probability is low enough to fulfil additivity) with a probability which depends on the rate at which the described phenomenon occurs. The pump rate $N_P$ represents the number of pumping processes per time step, where the latter is chosen small enough to ensure an outcome either $0$ or $1$, for most steps, to guarantee additivity. The values of the rates used in the simulations match the estimates for small nanopillar devices and amount to (cf.~\cite{Puccioni2015}): $\gamma_{\parallel} = 3 \times 10^9 s^{-1}$ for the spontaneous emission (process $N_d$), $\Gamma_c = 1 \times 10^{11} s^{-1}$ for the losses of the on-axis photons (i.e., stimulated, but also fraction of spontaneous photons in the lasing mode -- $L_S$ and $L_L$ processes), $\Gamma_o = 5 \times 10^{13} s^{-1}$ for the lifetime of off-axis (spontaneous) photons ($L_o$ process). $E_S$ is proportional to the product $\gamma_{\parallel} \beta S N$, as the standard probability of obtaining a stimulated process. 

The length of feedback is fixed at $L_{ext} = 60 cm$. The fraction $F$ of photons reinjected into the laser, and $\beta$ -- fraction of spontaneous emission coupled into the lasing mode --, are used as parameters in the simulations.

\subsection{Photon statistics}
The photon statistics of laser emission is mainly characterized through the second-order correlation function $g^{(2)}$, the most widely used indicator for the analysis of the statistical properties of light sources, defined, for a photon number $S(t)$ exiting the laser, by~\cite{Foster1998}

\begin{eqnarray}
\label{defg2}
g^{(2)} (\tau) & = & \frac{\langle S(t)S(t+\tau) \rangle}{\langle S(t) \rangle ^2} \, 
\end{eqnarray}     

\noindent where $S(t) = L_S + L_L$ is the total photon number exiting the cavity and $\langle S(t) \rangle$ indicates the time-averaged photon number emitted by the laser. Therefore, $g^{(2)}(\tau)$ describes the correlation between two temporally separated photons with time delay $\tau = t_2 - t_1$ from one light source, as can be measured by a Hanbury-Brown and Twiss setup~\cite{Brown1956}. In conventional (large) lasers, when the input power passes through threshold, there is an abrupt transition from $g^{(2)}(0) = 2$ (thermal state, corresponding to fully spontaneous emission) to $g^{(2)}(0) = 1$ (coherent state, totally stimulated emission)~\cite{Wang2015, Jin1994, Ates2007}, a feature which allows us to identify the onset of lasing (threshold) ~\cite{Rice1994}.

\section{Threshold dynamics and autocorrelation-based sensing}
The sensing tool we propose uses the threshold dynamics of small lasers (with different $\beta$ factors) subject to varying feedback levels. The method correlates the laser behaviour with the amount of feedback it receives, thus enabling the collection of information coming from the environment. We thus define the feedback fraction parameter:
  
\begin{eqnarray}
F = \frac{S_{in}}{S_{out}} \times 100\%
\end{eqnarray}

\noindent where $S_{in}$ is the number of photons that are actually coupled into the lasing mode, and $S_{out}$ is the photon number coupled out of the cavity. The normalized pump refers to the so called threshold pump $P_{th}$, given by $P_{th} = \frac{\Gamma_c}{\beta}$, which corresponds to the midpoint in the steep portion of the steady-state curve (cf., e.g., Fig. 1) representing photon number versus pump~\cite{Wang2020}.

One assumption is made when using this model: the interference between intracavity photons and reinjected one can be neglected. This holds when either (or both, of course) of these conditions are fulfilled: 1. the coherence length of the emitted radiation is smaller than the round-trip ($2 L_{ext}$); 2. the reflected component does not maintain a phase relationship (i.e., the reflection is not specular).
 
The first condition is rather easily fulfilled, thanks to the fact that we are using the threshold region, where the laser output is mainly composed of sharp pulses~\cite{Wang2015, Wang2019}. From the typical pulse duration ($< 1 ns$) we can extrapolate a minimum allowed length for the feedback arm $\approx 10 cm$ (simulations are carried out with a feedback arm six times longer than this limit).  The second condition is also not difficult to satisfy, since most objects are rough scatterers, on the wavelength scale, and will destroy the phase coherence of the reflected wave.  Thus, use of an photon-number-based model, eqs. (1-4), is fully justified.

\subsection{Threshold dynamics}
\begin{figure}[!t]
\centering
  \includegraphics[width=1.68in,height=1.34in]{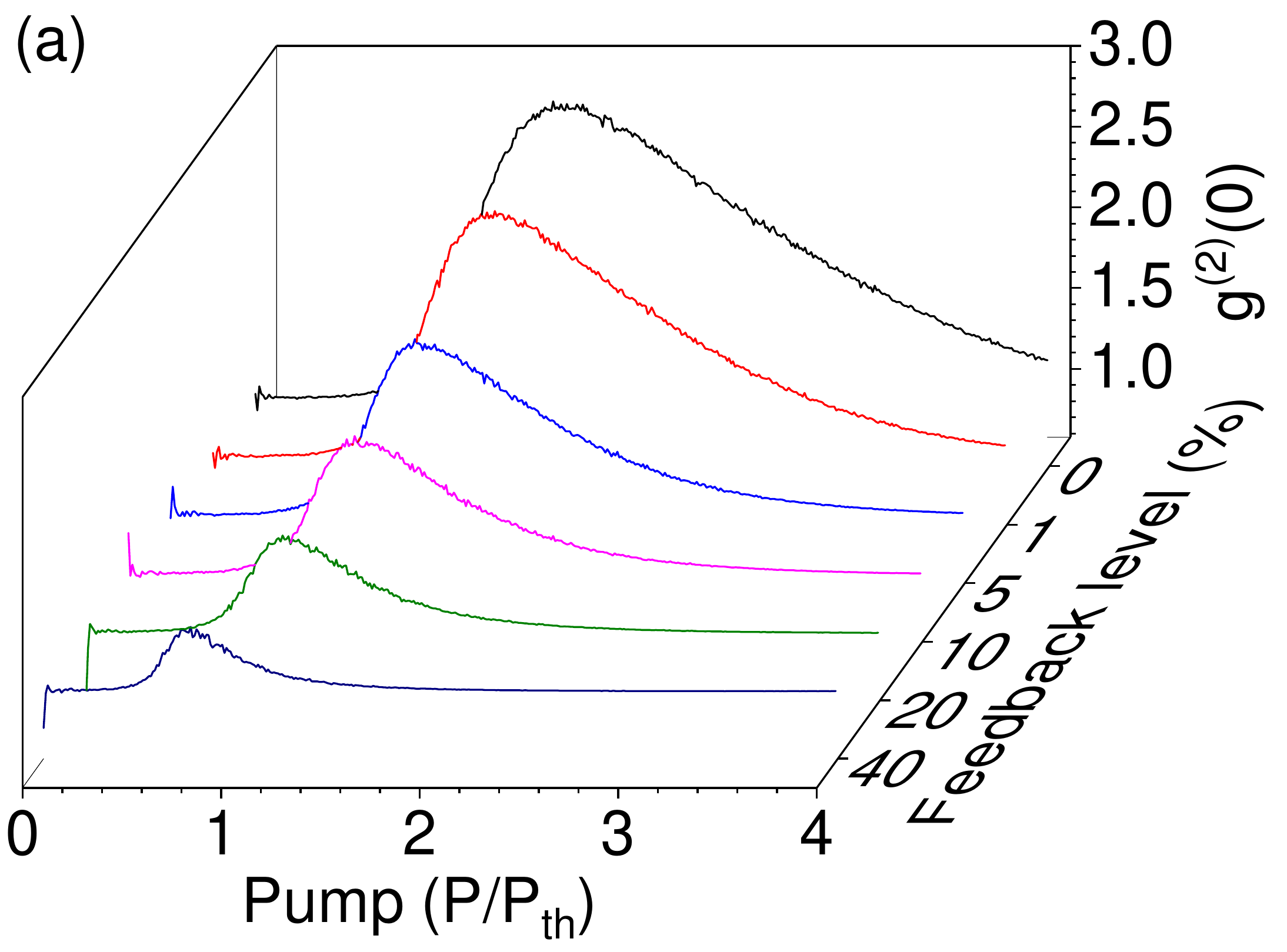}
  \includegraphics[width=1.68in,height=1.34in]{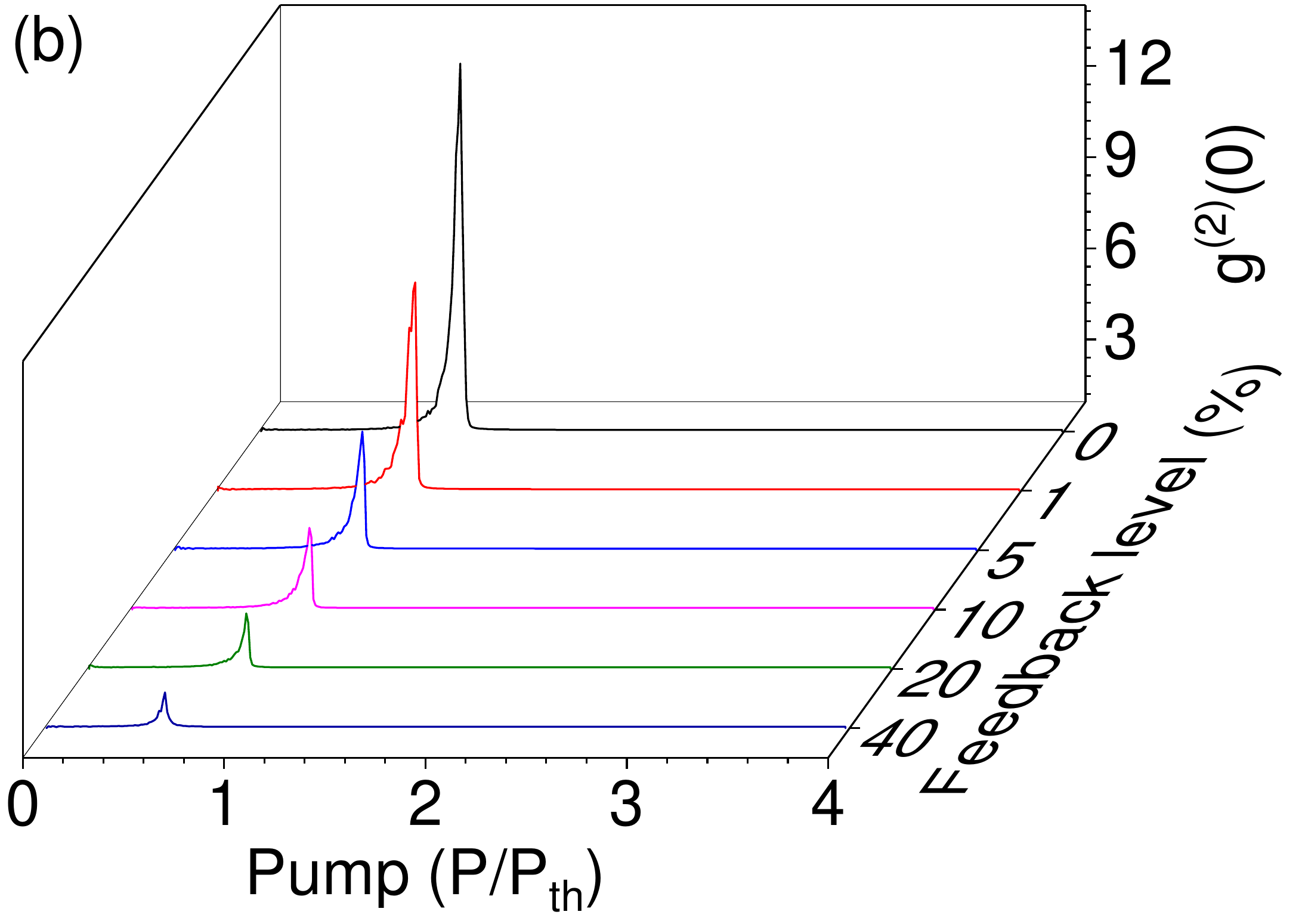}
  \caption{Second-order autocorrelation function spectra of nanolaser (a) and microlaser (b) under different feedback levels: 0 (black), 1$\%$ (red), 5$\%$ (blue), 10$\%$ (pink), 20$\%$ (green) and 40$\%$ (dark blue).}
  \label{AutoCorre}
\end{figure}

The statistical properties of the radiation emitted by the two laser kinds is analysed with the help of the second-order, zero-delay autocorrelation function, $g^{(2)}(0)$ (Fig.~\ref{AutoCorre}). For a nanolaser (Fig.~\ref{AutoCorre}a) the autocorrelation clearly shows superthermal emission appearing in the smooth transition between spontaneous and coherent regimes for $F < 10\%$.  The strongest superthermal emission appears in the absence of feedback since the latter acts as an external source which enhances the probability of starting a stimulated emission burst~\cite{Wang2019}. The more frequent bursts also reduce the excursions in population inversion fluctuations, reducing the amplitude of the photon bursts. The combination of more frequent and smaller burst gives rise to a monotonic decrement of the autocorrelation.

Accompanying the reduction in autocorrelation, we observe a shift in its peak towards lower pump values. This is consistent with the shift for the laser response where the fraction of photons reinjected into the cavity anticipates the transition towards lasing relative to the free-running operation.

The autocorrelation for the microlaser (Fig.~\ref{AutoCorre}b) shows an entirely different functional dependence on pump. Rather than a broad response, which matches the smooth growth of laser power as a function of pump in a nanolaser, we observe a sharp peak which grows rapidly and drops down abruptly towards coherent emission as a function of pump. As remarked for the nanolaser, the autocorrelation peak is gradually reduced as the feedback fraction increases, for the same reasons already discussed. The autocorrelation's peak shift to lower pump also matches both the laser characteristics for growing feedback and the observation in the nanodevice. What distinguishes, instead, the microlaser is the very large value of $g^{(2)}(0)$, much larger than in its nanosized counterpart.  

\begin{figure*}[!t]
\centering
  \includegraphics[width=6.5in,height=3.0in]{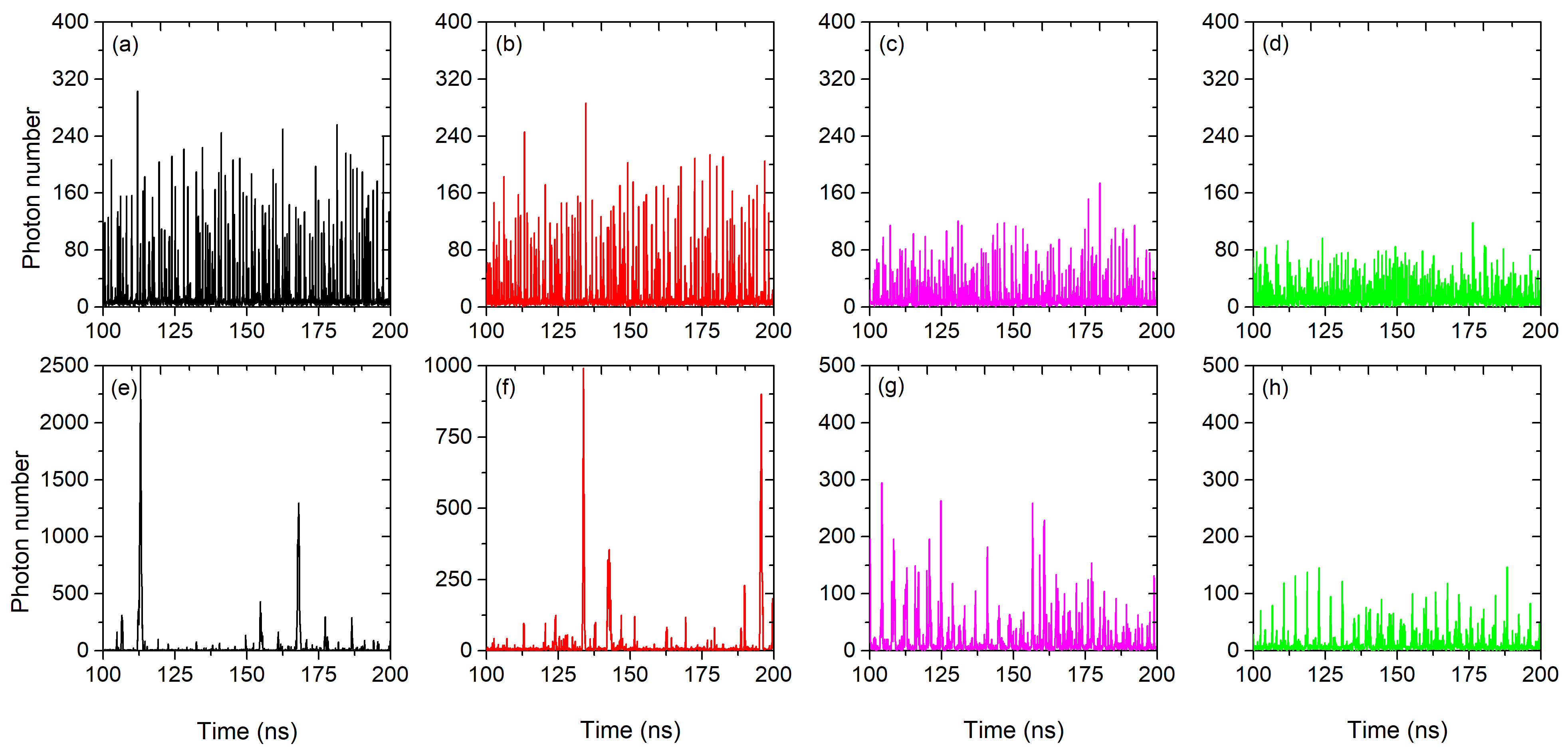}
  \caption{Temporal dynamics within 200 ns for the  nano- and microlasers with different feedback levels: (a) - (d) for nanolaser; (e) - (h) for the microlaser. For the feedback levels, (a) and (e) 0; (b) and (f), 1$\%$; (c) and (g) 10$\%$; (d) and (h) 20$\%$. }
  \label{TDynamics}
\end{figure*}

Further insight is gained from the temporal dynamics (Fig.~\ref{TDynamics}). The strongest pulses are observed for the free-running operation (panels (a), nanolaser, and (e), microlaser). As the feedback level increases, the pulses become more frequent and smaller, as already mentioned.  The consequence of the more frequent pulses is a progressive reduction in the value of $g^{(2)}(0)$. It is important to notice the different peak values for the two kinds of lasers, which contribute to the difference in peak autocorrelation value.  

\begin{figure*}[!t]
\centering
  \includegraphics[width=6.50in, height=3.0in]{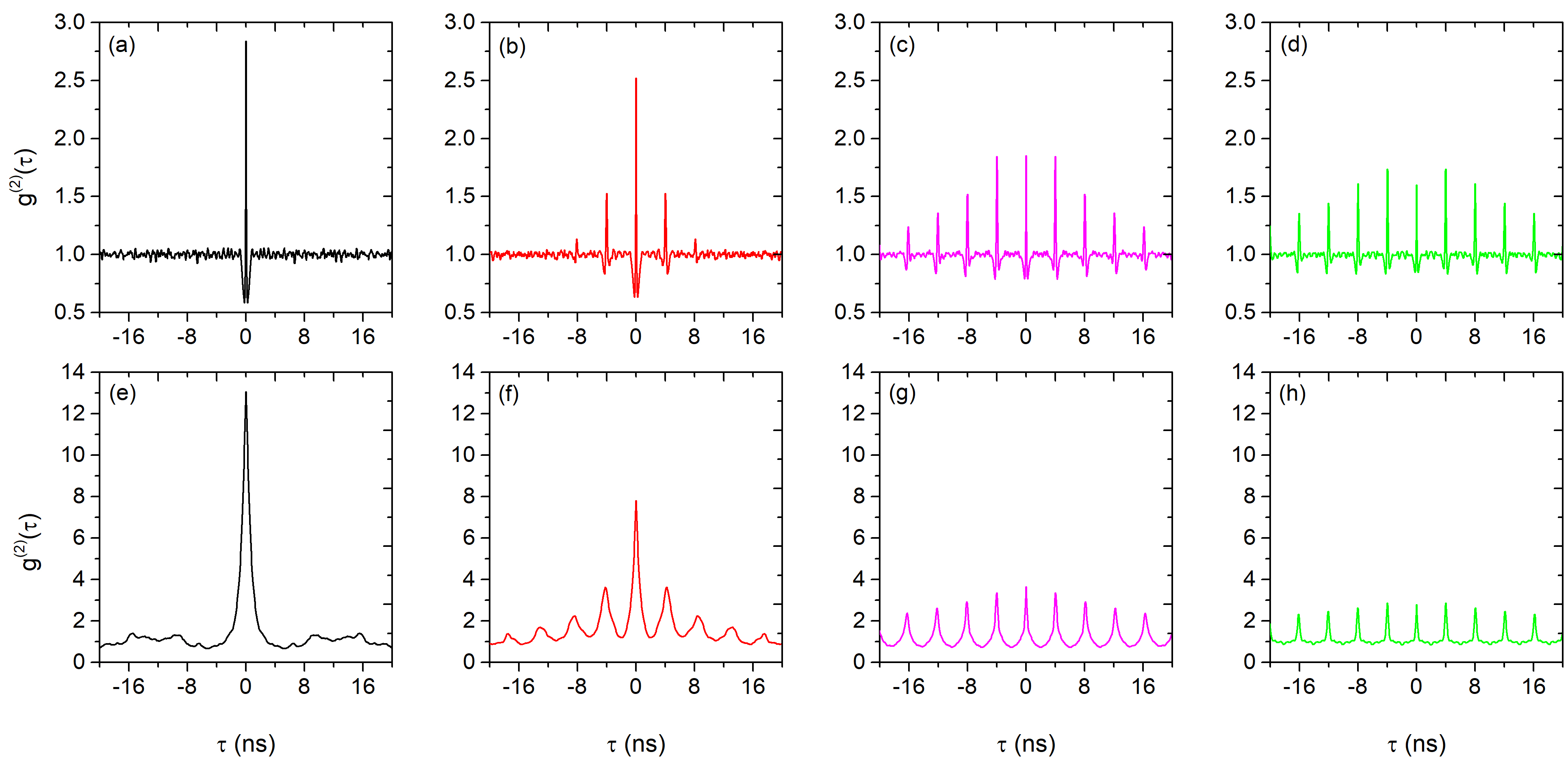}
  \caption{Second-order correlation traces in time delay (g$^{(2)}(\tau)$) of nanolaser (a-d) and microlaser (e-h): (a) and (e) 0; (b) and (f), 1$\%$; (c) and (g) 10$\%$; (d) and (h) 20$\%$. For nanolaser (a-d), the normalized pump values are 1.50, 1.42, 1.14, and 0.99; for microlaser (e-h), the corresponding pump values are 1, 0.99, 0.89, and 0.79.}
  \label{corretime}
\end{figure*}

Additional information is gained from examining the time-delayed, second-order autocorrelation. Fig.~\ref{corretime} shows the corresponding $g^{(2)}(\tau)$ for different feedback levels. For the nanolaser we observe sharp autocorrelation peaks, matching the narrow temporal spikes of Fig.~\ref{TDynamics}(a-d). A sharp anticorrelation ($g^{(2)}(\tau) < 1$) signals the fact that the peaks are well isolated from one another, independently of feedback level. Only for the largest amount of feedback that we examine here we observe a slightly more complex structure, which matches the evolution of the temporal signal into a denser sequence of pulses which, at times, broaden (Fig.~\ref{TDynamics}d). The anticorrelation feature surrounding each peak in $g^{(2)}(\tau)$ survives, though, even though it weakens in correspondence to the partial broadening of the peaks.  

The growth of side peaks, starting from the lowest feedback level considered, indicates the influence of feedback. Since in the pump regions we are considering the laser outputs photon bursts (Fig.~\ref{TDynamics}), a returning burst (even very small) induces a certain amount of correlation, highlighted by the presence of a peak in $g^{(2)}(\tau)$ at the delay time which corresponds to one roundtrip (e.g., Fig.~\ref{corretime}b) -- the peak is quite small for the second rountrip, implying that there is little correlation between pulses which are two roundtrips distant. For larger feedback levels, the correlation between more distant peaks grows and with it the number and height of the peaks that appear on the sides of the autocorrelation. Of course, the physical mechanism is based on the stimulation of a pulse by an arriving one (fed back from the scatterer) and in setting up a chain of such pulses whose efficiency grows larger, thus establishing long range correlations in the form of a ``comb'' which slowly decays away. Such comb is not easily recognized from the temporal trace (Fig.~\ref{TDynamics}c,d) due to residual irregularities in amplitude and repetition, but is clearly identified by the time-delayed autocorrelation. Interestingly for the largest feedback level examined (Fig.~\ref{TDynamics}d) the central peak is smaller than the first recurrence, suggesting that the dominating mechanism for photon burst generation is no longer a spontaneous fluctuation~\cite{Wang2019}, but, rather, the external self-injection of a pre-existing pulse. Finally, it is important to notice that the long-term coherence which emerges from the structure of $g^{(2)}(\tau)$ is not induced by phase coherence in the signal, but only by the more or less regular repetition of photon bursts, each with a low degree of coherence. 

The microlaser is characterized first of all by a very large value of the autocorrelation at $\tau = 0$, much larger than what was found for the nanolaser. The origin of this extreme superthermal emission is visible in the temporal sequence: very large and widely spaced photon bursts. This extreme behaviour, typical of the larger device, originates from the strong decoupling between population and photon fluctuations below the threshold for continuous emission~\cite{Wang2019}.

The structure of the time-delayed autocorrelation is characterized by a central peak, much broader than in the nanolaser, with an irregular structure as a function of delay time (Fig.~\ref{corretime}e). This can be easily understood again from the temporal sequence:  the photon burst does not consist of a simple, narrow spike, but can have a complex structure (with multiple peaks) and is considerably broader. This more convoluted pulse conformation also explains the appearance of shallower anticorrelation regimes over broader time delays, as there are time intervals in which a second pulse is inhibited (Fig.~\ref{corretime}e).

Feedback does not substantially change the peak width in $g^{(2)}(\tau)$, but substantially lowers the amplitude of the central peak.  Indeed, a returning photon burst appears a bit more effective in generating a new pulse than in the nanolaser, and in particular in stringing along a chain of bursts:  even with only 1\% feedback level a correlation immediately emerges even at the level of 5 external cavity roundtrips (vs. 2 in the nanodevice).  The main effect of increasing feedback is an equalization in the time-delayed peaks:  the central on is progressively reduced due to the reduction in peak amplitude and by the progressive increase in the fraction of time occupied by the peaks; the side-peaks instead gain in relative strength as a pulse comb becomes more and more apparent (even though not too easily recognized in the temporal domain, Fig.~\ref{TDynamics}g,h).     
 
\subsection{Sensing implementation}\label{simp}
\begin{figure*}[!t]
\centering
  \includegraphics[width=1.8in,height=1.5in]{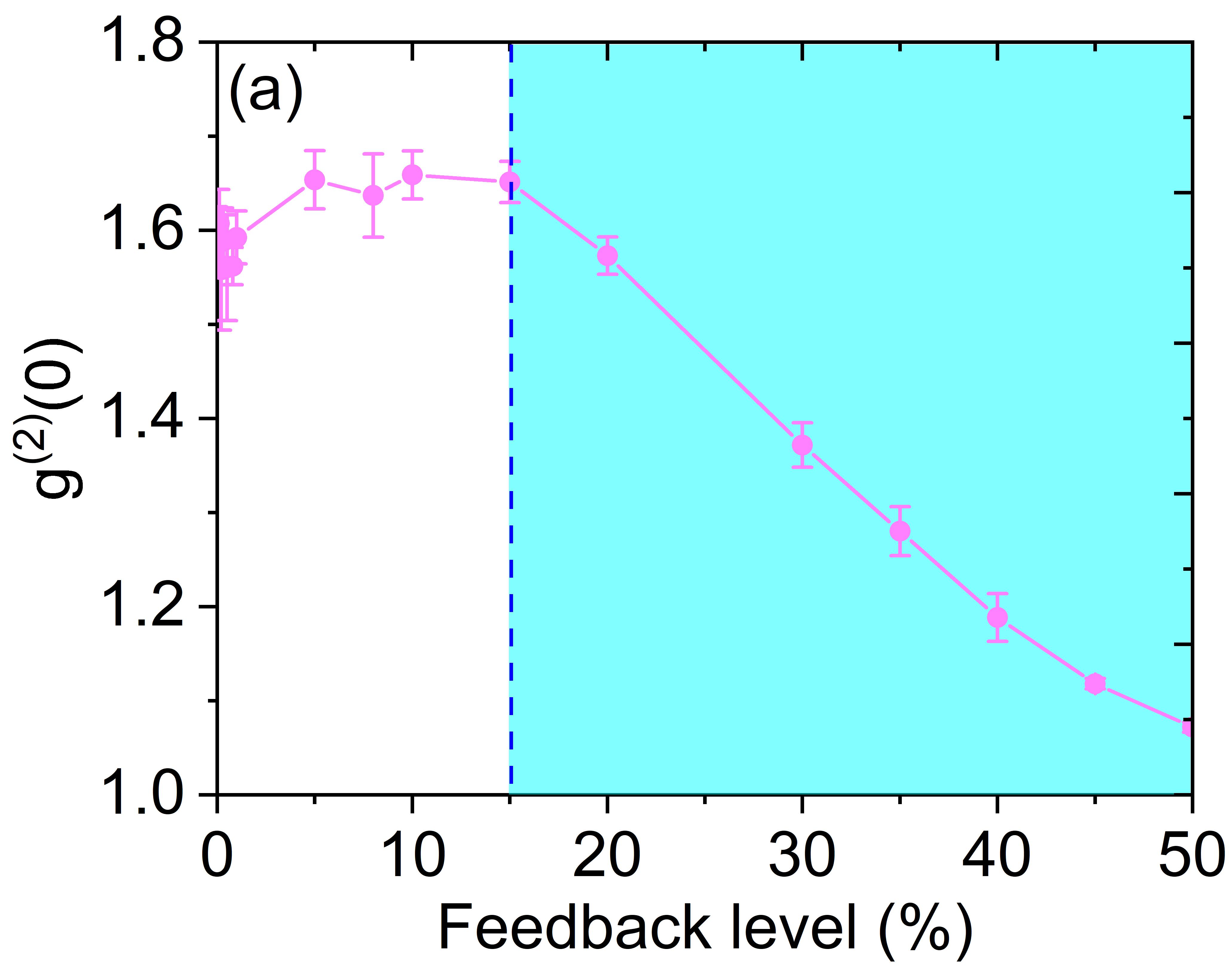}
  \includegraphics[width=1.8in,height=1.5in]{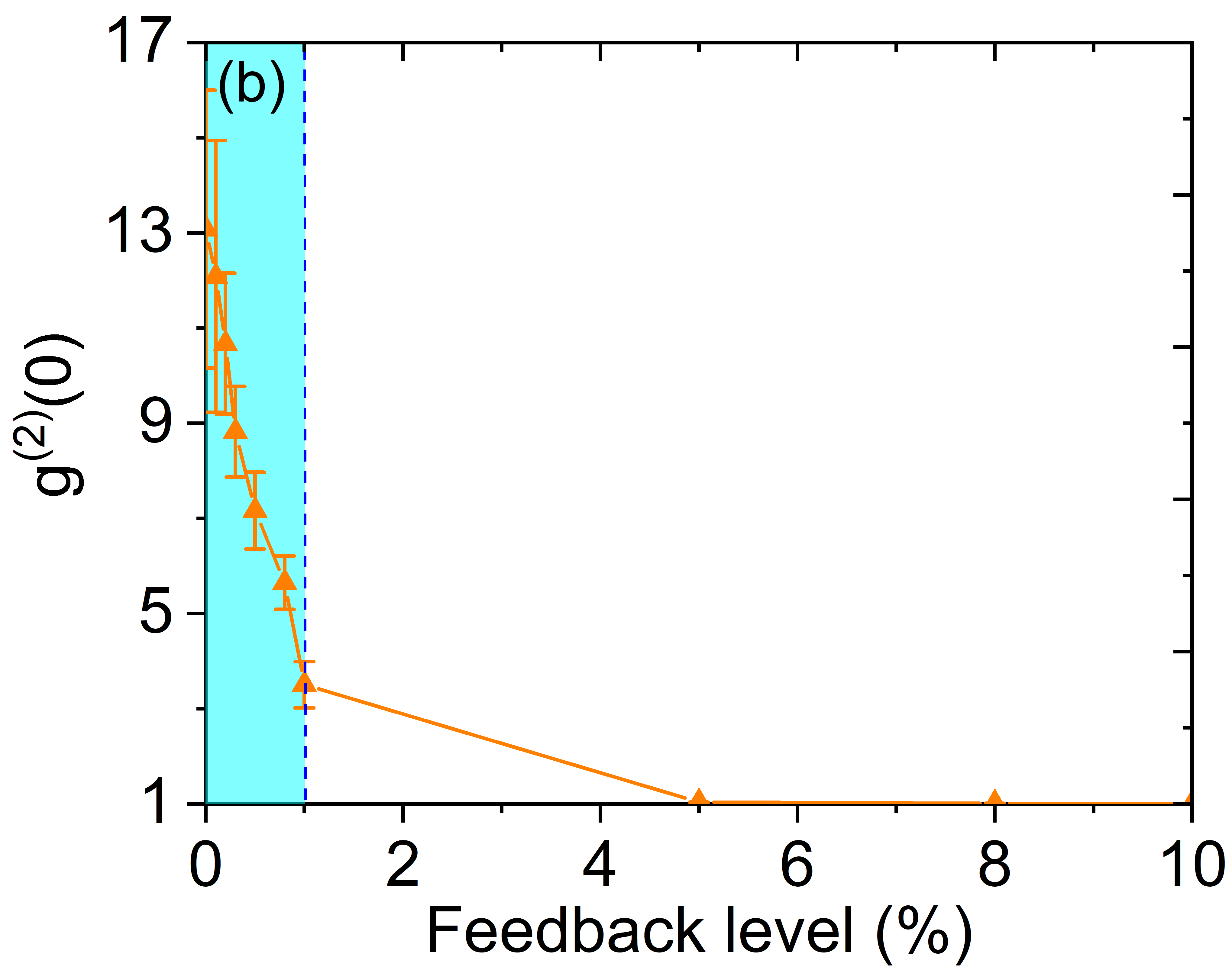}
  \includegraphics[width=2.0in,height=1.5in]{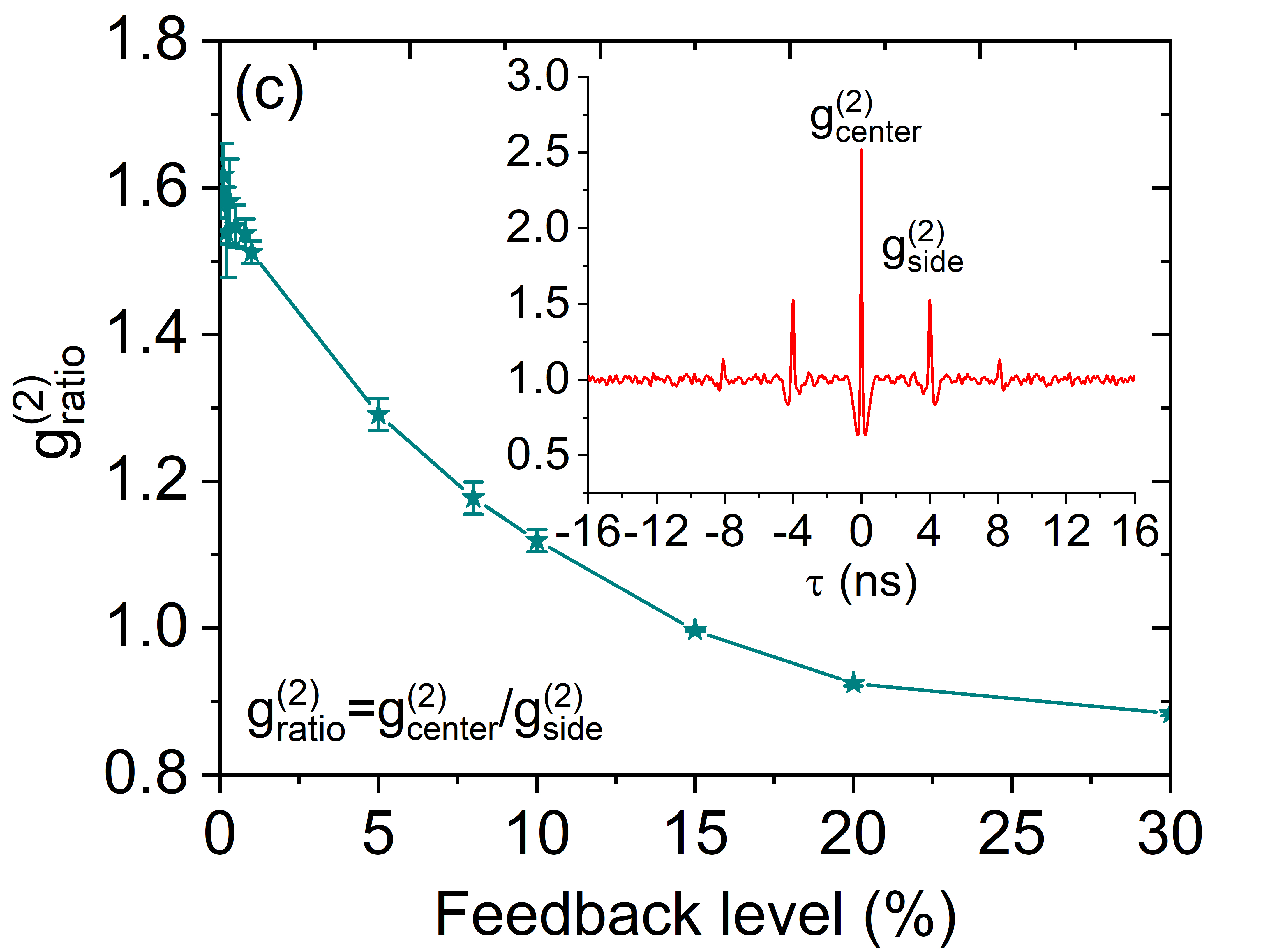}
  \caption{g$^{(2)}(0)$ and g$^{(2)}_{(ratio)}$ responses of small lasers at P = P$_{th}$ to the external feedback: (a) g$^{(2)}(0)$ response of nanolaser; (b) g$^{(2)}(0)$ response of microlaser; (c) g$^{(2)}_{(ratio)}$ response of nanolaser, inset, temporal second-order correlation function curve. g$^{(2)}_{(ratio)}$ indicates the ratio between central peak and first side peak.}
  \label{auto}
\end{figure*}

Two sensing schemes emerge from these considerations. The first one is based on the measurement of the zero-delay autocorrelation ($g^{(2)}(0)$) and exploits the complementary performance of the two laser groups (nano- vs. microlaser). Fixing the pump at its threshold value ($P = P_{th}$), Fig.\ref{auto}a shows that low levels of feedback (up to $\approx 15$\%) in the nanolaser lead to a flat response in $g^{(2)}(0)$, thus no discrimination. Instead, for larger feedback levels (up to 50$\%$) there is a strong monotonic decrease in the autocorrelation (shadowed region), with fluctuations sufficiently small to provide good feedback level identification.  

The converse is true for the microlaser, which displays a sharp decrease for extremely weak feedback ($< 1\%$) with good sensitivity in this range (shadowed region in Fig.\ref{auto}b). Following the sharp drop, the microlaser response to feedback is no longer sensitive to its variations and provides no information. These two implementations of the first scheme enable therefore feedback detection in two disjoint intervals.


The second scheme makes use of the time-delayed second-order autocorrelation ($g^{(2)}(\tau)$) and of the appearance and structure of the revival peaks. Introducing as an indicator the ratio between the central ($\tau = 0$) and first side peak ($\tau = 2 \tau_{ext}$), we exploit the relative evolution of the autocorrelation structure to extract information on feedback. Fig.\ref{auto}c shows the ratio (defined in the inset) as a function of feedback for the nanolaser. The indicator clearly shows that this choice neatly fills the gap in the feedback strength left by the previous technique, satisfactorily covering the interval $1\% < F < 20\%$.  This scheme therefore completes the coverage of the feedback range (as indicated by Fig.~\ref{corretime}e-h, the structure of the microlaser autocorrelation does not lend itself for a similar treatment).

Summarizing the results of this section, we find that microlasers are best suited for measurements in the low feedback range (using $g^{(2)}(0)$; $F \lessapprox 1$\%).  The sharp dependence of the zero-delay autocorrelation on the amount of feedback renders them extremely suited to discern tiny amounts of reinjected light.  We can thus consider the microlaser as an extremely sensitive feedback detector.  The nanolaser, on the other hand, offers a broader range of reinjected power levels using two complementary techniques.  For $0.01 \lessapprox F \lessapprox 0.2$ the ratio between the central and first side-peak in the autocorrelation (Fig.\ref{auto}c) is an effective indicator, while for $0.15 \lessapprox F \lessapprox 0.5$ the zero-delay autocorrelation is preferable. 

The failure for the microlaser in providing usable information with the peak ratio scheme is intrinsically related to its superior performance at very low reinjection levels; indeed, the strong nonlinear amplification of the (re-)injection seeding leads on the one hand to the best sensitivity at low power, but also to larger pulse fluctuations.  When the differential sensitivity (slope of $g^{(2)}(0)$) is large, the fluctuations are offset by the strong changes (cf. Fig.~\ref{auto}b); however, they may mask the changes in feedback amplitude for a smoother response ($F \gtrapprox 0.01$).  Conversely, the nanolaser, with its reduced sensitivity, offers much better performances in the feedback range where smaller error bars are required and promises to be a valuable device in this range.

Before concluding, it is worth pointing out a proposal to obtain strong light bunching from nanodevices~\cite{Jahnke2016}.  Experimental implementations of such scheme may offer, in the future, interesting applications -- e.g., in sensing -- due to the much larger amount of light bunching thus obtainable from a nanolaser.

\section{Sensing considerations}
We now examine the main features of a sensor based on the detection properties described in the previous sections. In a first subsection (section IVA) we discuss some details of an experimental implementation of the proposed scheme, based onto two different kinds of realization.  In addition, two alternative detection strategies are explicitly discussed. The flow configuration is one that offers versatility, but is not a requirement of the scheme; a device could also operate in a static configuration (e.g., scanning a sample in a sequential way to reconstruct a matrix of measurements), thus replacing the flow speed considerations with the measurement time at each spot. We further identify the range of particles sizes which can be detected as a function of their {\it effective surface}, determined on the basis of geometrical considerations scattering amplitude (Section IVB).

Section IVC discusses the features of the two detection schemes that we propose in relationship with the flow speed limitations and data acquisition rates. The flow speed limitations are explicitly examined here. Finally, Section IVD analyzes the possibility for, we discuss the possibility for counting multiple particles simultaneously, either in a monodisperse sample or in a polydisperse one. Of course, the integral information is always available and provides a measure of the total amount of scatterers present in the sample.

As will become clear in the following, the information coming by the sensing scheme combines different scatterer properties, such as surface and scattering coefficient, whence the introduction of the effective scatterer surface.  Scattering coefficients can largely vary depending on impinging light wavelength and material properties; in addition, part of the radiation may be absorbed. Our sensing scheme is based on ultralow average power, ranging from sub-nW to $\mu$W, thus the temperature increase to be expected from the scatterer is negligibly small over the interaction time (cf. residence times in the beam, estimated below).  The fluid in which the particles move, or are advected, could also absorb part of the radiation, but the same considerations hold, as they do for the particles.  The speed at which scatterers more through the beam, in the flow configuration, is sufficiently slow (typically 1 m/s) to exclude dynamical scattering effects which may affect the measurement.  When scanning a static sample, instead, the measurement is taken with the beam at rest on the desired position, thus excluding unwanted consequences.  Finally, an overall calibration coefficient -- taking into account the geometrical factors related to the fraction of feedback truly entering the laser -- can be added to the quantitative considerations on the basis of instrumental tests.

\subsection{Experimental implementation}\label{expt}
\begin{figure}
\centering
  \includegraphics[width=3.2in]{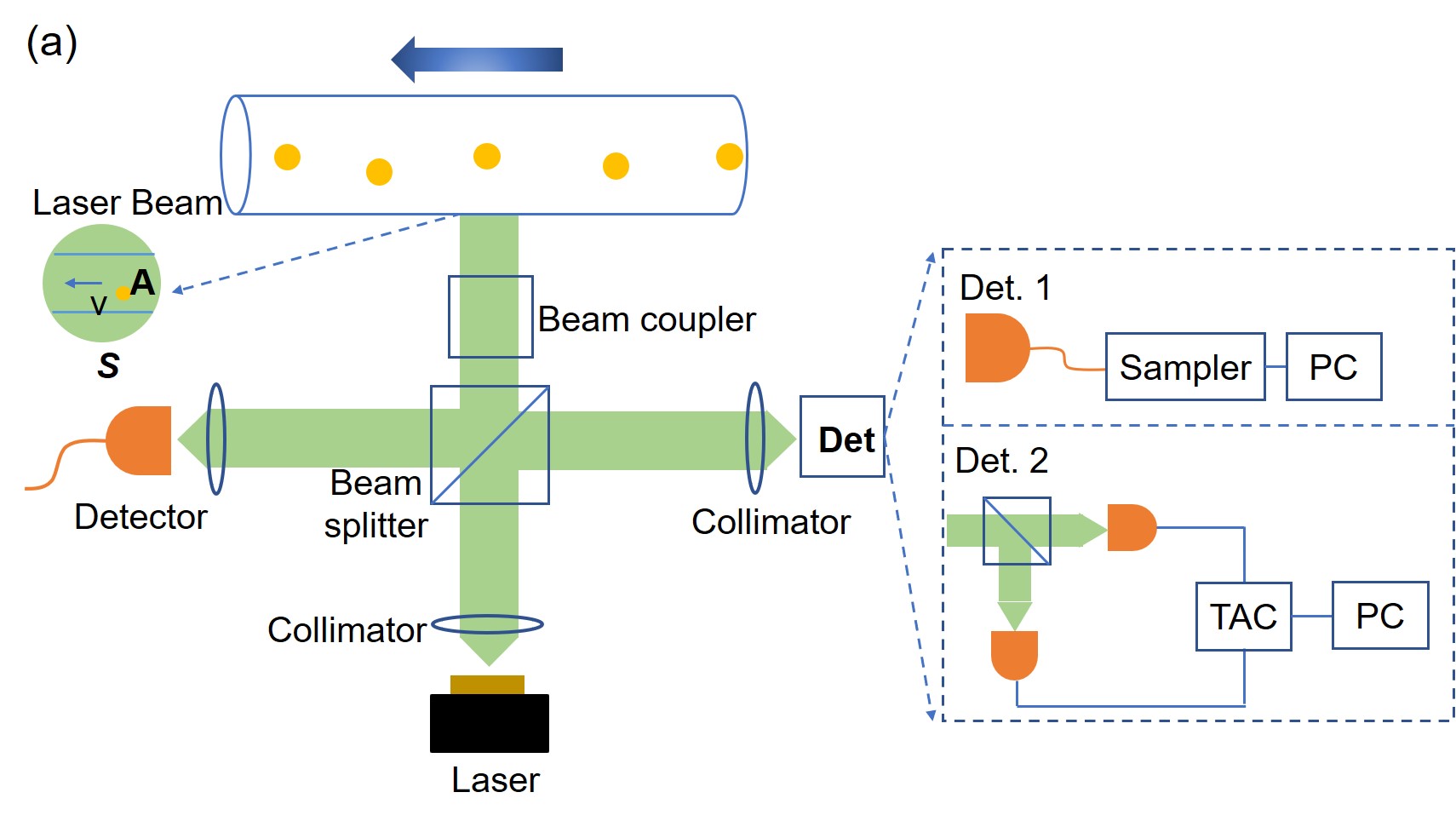}
  \includegraphics[width=3.2in]{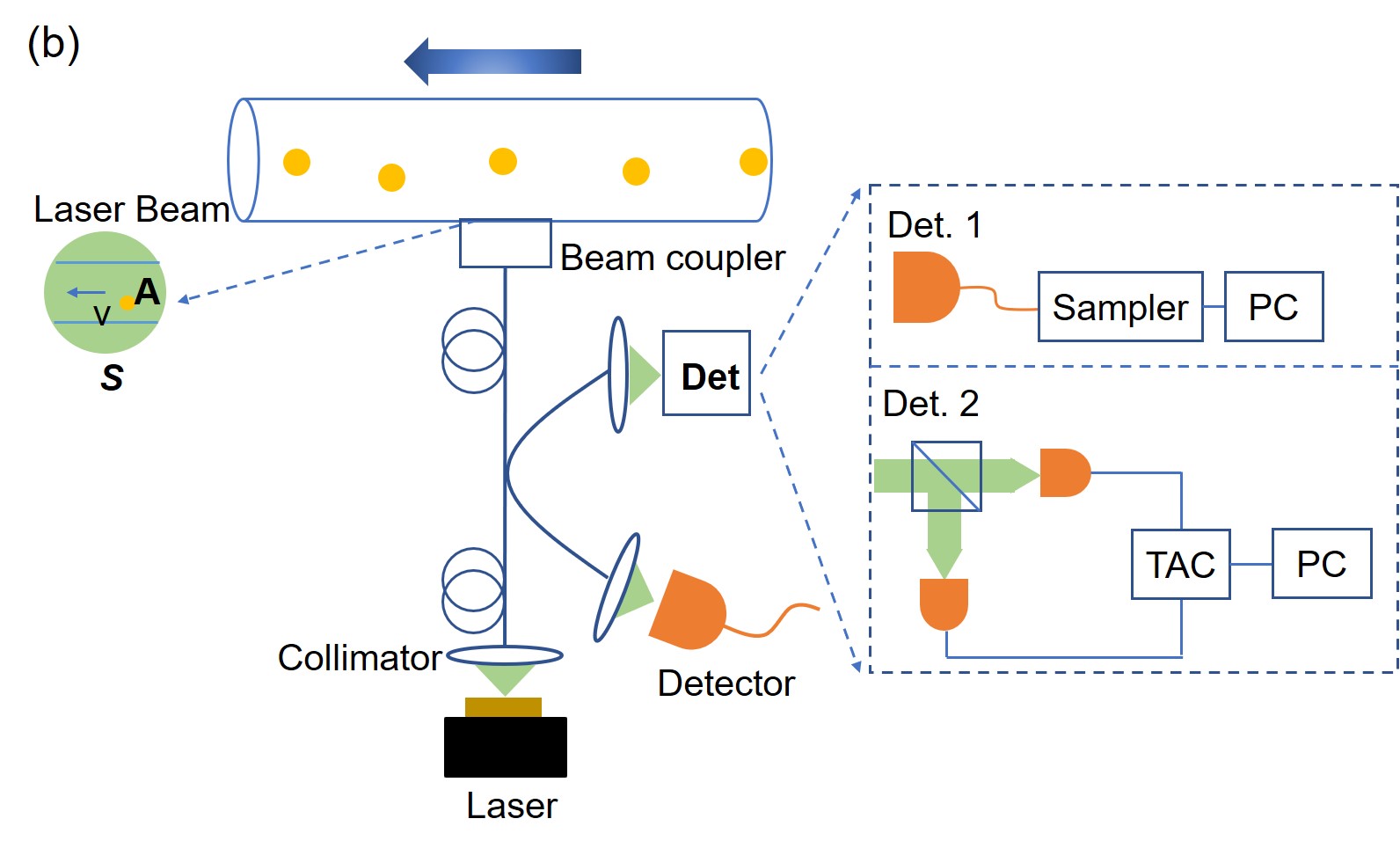}
  \caption{Two schematic diagrams of experimental design realized with discrete components (a) or in a fibered configuration (b).  Particles move at speed $v$ inside the flow-cell (top of diagrams) from right to left.  $A$ represents the scatterer's cross-section while $S$ denotes the laser beam's surface.  The laser output goes through a collimation stage (a) (not needed in the fibered laser) and is split (either with a discrete or a fibered component).  The component travelling towards the experimental cell is matched by a Beam coupler (both setups) to select the laser surface $S$.  The backreflected light returns towards the laser and can be monitored by the secondary port of the splitter (lens-coupled detector).  The light reaching the laser feeds the signal back into it and initiates the dynamics, whose temporal behaviour is monitored by the main splitter port.  For both setups ((a) and (b)) two different detection schemes are possible:  linear detection (Det. 1) or photon counting (Det. 2).  The former is suited to microlasers and can be realized with a single detector connected to a digitizing sampler and a PC.  The latter is best obtained with a HBT setup where two photon counting detectors, connected to a TAC,  enable the real-time reconstruction of the statistics with a PC.  Details in the text.}
  \label{setup}
\end{figure}

The experimental implementation of our scheme can take different forms.  In this subsection we illustrate a few key points of the setup.   
A schematic of principle for the experimental design is shown in Fig.~\ref{setup} (with discrete components (a) or in fibered form (b)): the light emitted by a small laser, after collimation, separation and matching to the experimental sample, impinges onto the particle (orange dot) contained in the measurement cell.  The backscattered component of the light goes through the inverse path and enters the laser again, thus stimulating the temporal dynamics of the emitted light intensity.  The collimator is (normally) needed for very small lasers, due to their intrinsically large divergence (for the fibered scheme, the collimating action can be played by an integrated optics coupling between laser and fiber).  The two-port beam splitting stage is needed for signal analysis, while the matching step (Beam Coupler) offers additional flexibility in adapting the beam size to the sample's.  This last step is not indispensable, though, and can be left out in case space or cost considerations are an issue. 

The laser beam in the measurement area is represented by a green circle, of surface $S$ (cf. side sketch), with one particle flowing through it with velocity $\vec{v}$. The scatterers are assumed to flow through the beam one at a time (low density) for a first measurement configuration (cf. section~\ref{manyp} for considerations about measurements at larger densities).  As already mentioned, the Beam Coupler ensures proper light coupling to the cell, control of the beam size $S$ and collection of the scattered light for reinjection into the laser. Polarizing elements (not shown) can be added both to the laser emission and to the detection branches for further refinements.

The beam splitting leads to the potential use of two separate detection units. The port which directly views the light backscattered by the sample may be used for signal monitoring, but is not central to our design. The main port (viewing the light issued by the laser, thus monitoring the dynamics) can be equipped with either of two detection units.  

The first one (Det. 1) is based on linear detection using, for instance, a fast, very sensitive photodetector (e.g., Thorlabs PDA8GS).  The electrically converted signal is sampled and sent to a PC for the reconstruction of the autocorrelation function (cf. below for details).  This scheme is suited to microlasers, since their power output is sufficiently large for this simpler and less expensive configuration.

The second one (Det. 2) makes use of two photon-counting detectors in a Hanbury-Brown and Twiss (HBT) start-stop configuration~\cite{Wang2020a}.  A Time-to-Amplitude Converter (TAC) sends to the PC the time delays from which the autocorrelation function can be recostructed.  This scheme, indispensable for nanolasers,  can also be used with microlasers.

\subsubsection{Real-time computation of the autocorrelation}

The two detector configurations have different requirements for the real-time computation of the autocorrelation.  Standard TAC data are automatically sorted by the accompanying software -- home-written or commercial -- to provide the full $g^{(2)}(\tau)$ function as the data are accumulated (e.g.~\cite{Wang2020a}).  It is therefore sufficient to extract the $\tau = 0$ component for $g^{(2)}(0)$ and perform a search for the first peak $g^{(2)}(\tau_1)$ ($\tau_1 = 2 \tau_{ext}$) to compute the ratio $g^{(2)}_{ratio}$ (Fig.\ref{auto}c). 

Besides providing the result immediately at the end of the measurement, this method has the advantage of enabling a decision on the duration of the latter through a test which can be performed as the data are accumulated.  Defining a generic variable
\begin{eqnarray}
G & = & 
\left\{
\begin{array}{c c l}
g^{(2)}(0) & : & {\rm zero \, delay \, autocorrelation}\\
g^{(2)}_{ratio} & : & {\rm side \, peak \, technique}\\
\end{array}
\right.
\end{eqnarray}
the acquisition can be stopped as soon as convergence is attained.  The latter can be monitored by setting the desired relative precision $G_r$ and checking
\begin{eqnarray}
\left| \frac{G_j - G_{j-1}}{G_{j-1}} \right| & \le & G_r \, ,
\end{eqnarray}
where the fulfillment of the inequality serves as a stopping signal.  The advantage is a gain in acquisition time and, therefore, a possible increase in instrumental performance.  In the following, estimates related to acquisition time are based on standard number of points which are typically sufficient for a good result and typically correspond to a worst-case scenario.  The automatic stopping described here can provide faster estimates.  Of course, a minimal number of samples needs to be accumulated before the test makes sense, but this is best left to field tests.

Data acquisition with scheme Det. 1 does not automatically provide the autocorrelation, which needs to be computed from eq.~(\ref{defg2}).  However, the fast performance of CPUs ensures the possibility for real time -- i.e., point by point -- computation of the autocorrelation with the following scheme:
\begin{eqnarray}
\label{incrav}
\langle S_j \rangle = \frac{j-1}{j} \langle S_{j-1} \rangle + \frac{S_j}{j} \, , \\
\label{incrav2}
\langle S^2_j \rangle = \frac{j-1}{j} \langle S^2_{j-1} \rangle + \frac{S^2_j}{j} \, ,
\end{eqnarray}
where $\langle S_j \rangle$ denotes the average obtained by adding the $j$-th point to the previous average $\langle S_{j-1} \rangle$, and similarly for the square of the signal $S$.  Substitution of the values into eq.~(\ref{defg2}) immediately provides an estimate of the zero-delay autocorrelation function up to the $j$-th point.  A trivial extension enables also the computation of the time-delayed version of the autocorrelation, as shown by a piece of sample code in C language (cf. Supplementary Material).  This step replaces the construction of the autocorrelation function in scheme Det. 2, while the rest of the procedure remains the same.   It is important to remark that the actual value of $g^{(2)}(0)$ is scaled by the ratio between the detection bandwidth and the laser's response time~\cite{Wang2020a}.  However, noise can be kept low enough to obtain good quality -- albeit rescaled -- values of the autocorrelation~\cite{Wang2015}.

We stress the fact that the Det. 1 scheme is presented as a possibly simpler and less expensive alternative, while remaining more limited in scope since it requires a larger amount of signal and is in principle more subject to noise.  The second scheme, instead, can fulfill all the sensing requirements.

\subsection{Detection ability}
The particle's amount of backscattered light is characterized by a proportionality coefficient $c$ which determines the amount of feedback. Here, we assume the rest of the optical setup to collect all the light, for simplicity (any losses on an actual implementation can be introduced as a multiplicative correction factor $c_s$, thus leading to a new backscattering coefficient $c^{\prime} = c \cdot c_s$). We further simplify the discussion, to illustrate the principle, by considering a {\it top hat} beam configuration~\cite{Siegman1986}. The feedback fraction parameter becomes therefore

\begin{eqnarray}
F & = & c \frac{A}{S} \, , 
\end{eqnarray}

\noindent where $A$ represents the particle's surface with scattering coefficient $c$.  This determination allows for the connection between {\it effective particle surface} $A_c = c A$ and feedback $F$, thus enabling the application of the concepts exposed in the previous sections. Notice that the beam surface $S$ can be adjusted with the coupling optics, provided some constraints which emerge later are fulfilled, thus lending a degree of versatility to the setup.

\begin{figure*}
\centering
  \includegraphics[width=1.85in,height=1.6in]{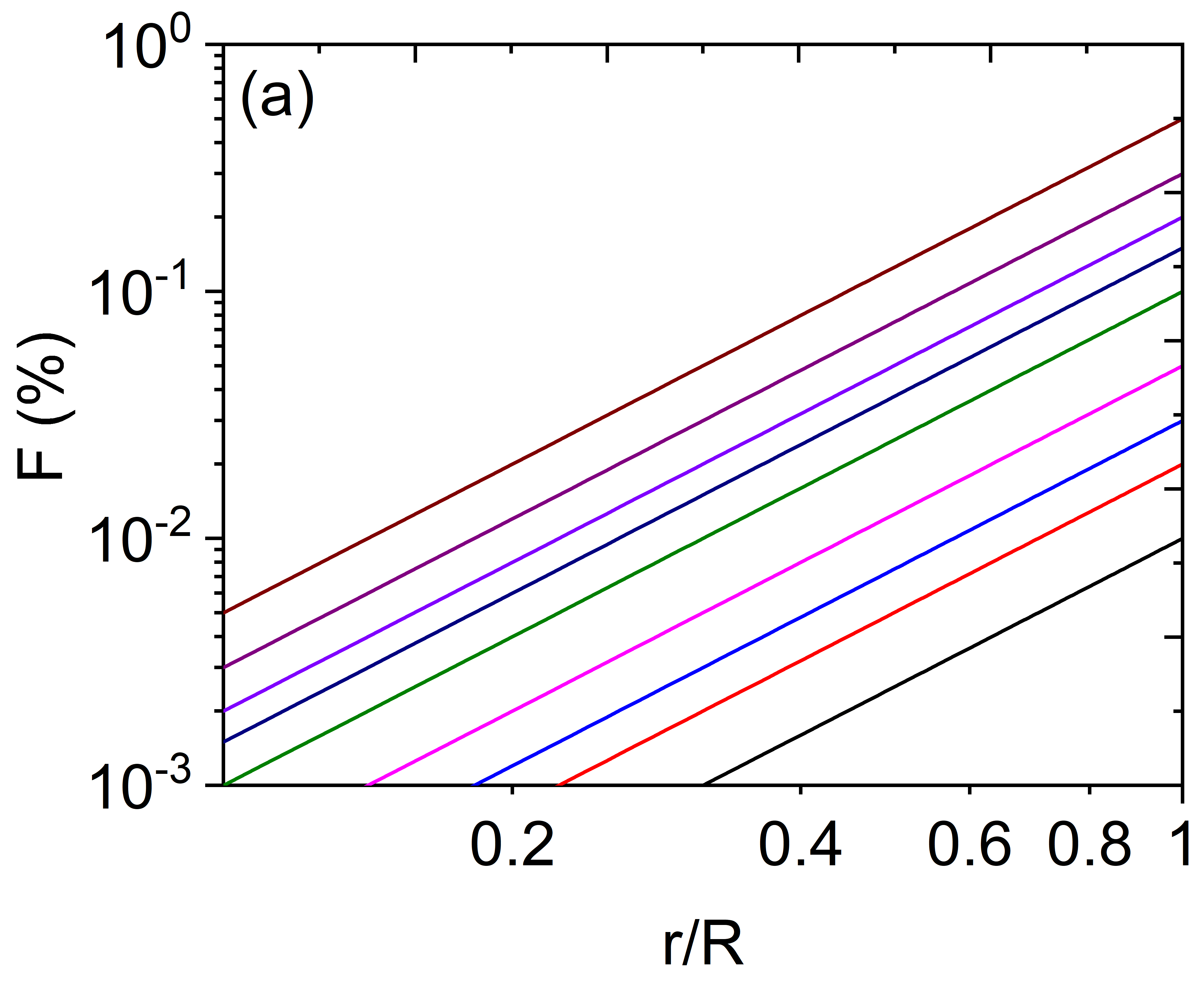}
  \includegraphics[width=1.85in,height=1.6in]{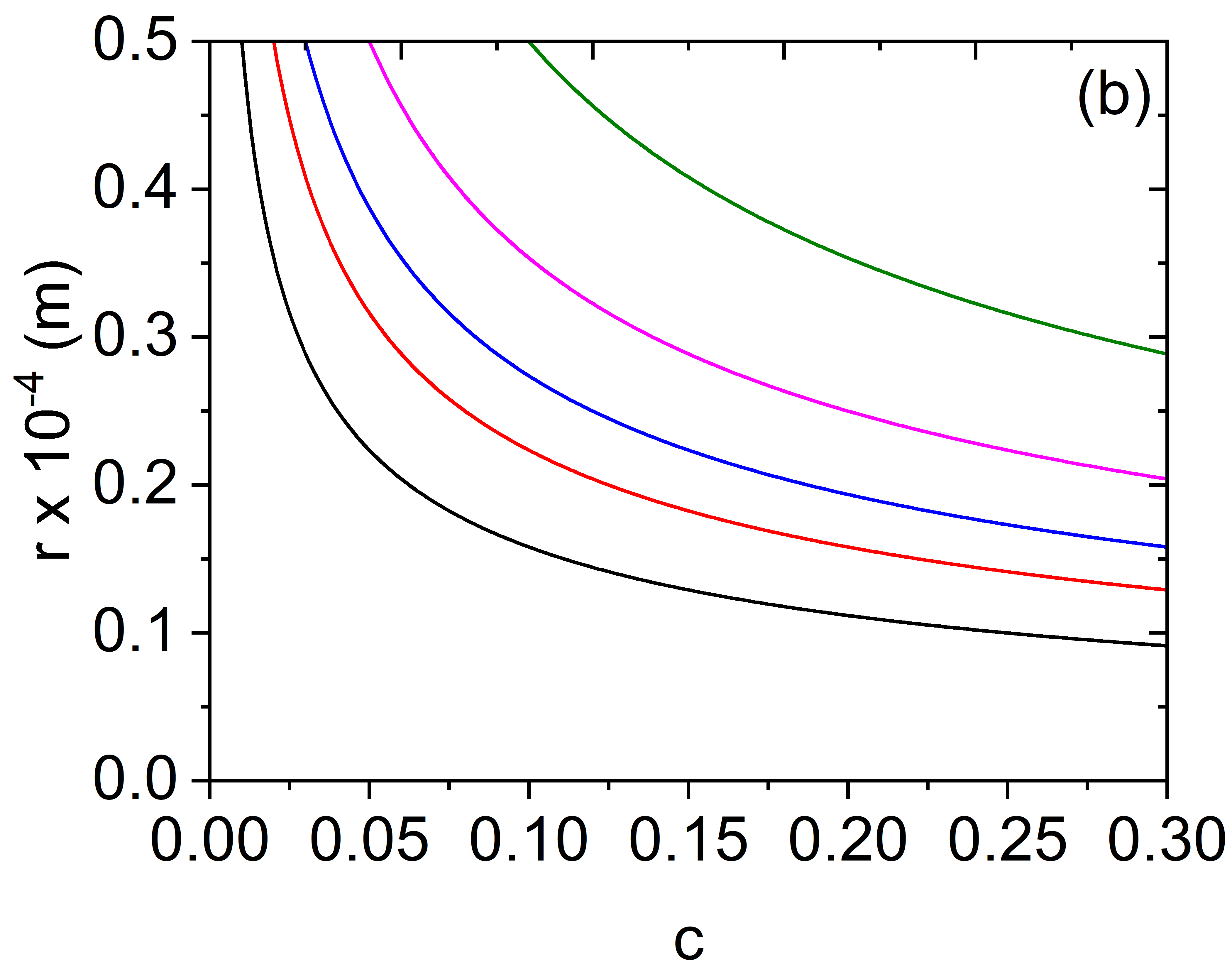}
  \includegraphics[width=1.90in,height=1.6in]{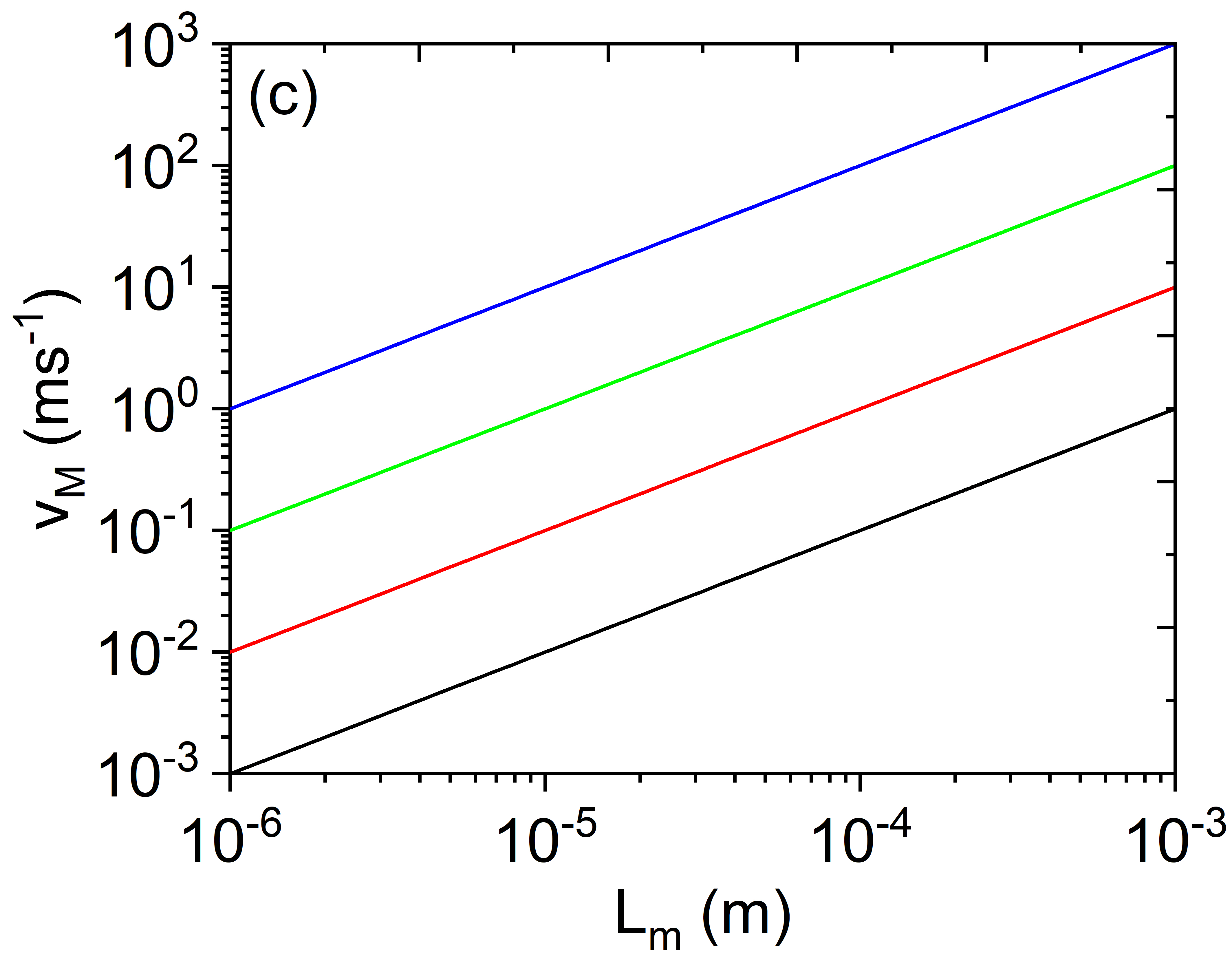}
  \caption{
  (a) Feedback fraction as a function of the particle radius r, measured to the beam radius in the cell R, for different values of the scattering coefficient c. From bottom (black) to the top (dark red), c = 0.01, 0.02, 0.03, 0.05, 0.10, 0.15, 0.20, 0.30 and 0.50, respectively; (b) detectable's particle radius r as a function of its scattering coefficient c for constant feedback fraction F. For this plot, the laser beam's radius in the cell is R = 50 $\mu$m. From bottom (black) to the top (green), F = 1$\%$, 2$\%$, 3$\%$, 5$\%$, and 10$\%$, respectively; (c) maximum allowed speed value for the particle, depending on the measurement path length L$_m$, for two different system bandwidths $\nu_B$ and number of measurements N$_m$. Black line, $\nu_B$ = 10$^8$ Hz and N$_m$ = 10$^5$; red line, $\nu_B$ = 10$^8$ Hz and N$_m$ = 10$^4$; green line, $\nu_B$ = 10$^{10}$ Hz and N$_m$ = 10$^5$; blue line, $\nu_B$ = 10$^{10}$ Hz and N$_m$ = 10$^4$. 
  }
  \label{fbck}
\end{figure*}

Fig.~\ref{fbck}a shows the amount of feedback that can be expected as a function of the relative radius $\frac{r}{R}$ ($r$ particle's radius and $R$ beam radius in the cell) for different values of the scattering coefficient. It is obvious that for particles with a low scattering coefficient the size of the beam has to rather closely match that of the particle, thus imposing constraints on the speed (cf. section~\ref{dets}). It is important to note that, if desired, the amount of feedback can be reduced to operate the sensor in its optimal range by enlarging the size of the beam relative to the particle's ($r/R$).

The exploitable feedback interval accessible for the determination of the particle's characteristic is delimited by the boundaries shown in Fig.~\ref{fbck}a, where we denote with $F_m$ and $F_M$ the minimum and maximum useable feedback values, respectively. Thus, particles whose effective size fulfills the conditions

\begin{eqnarray}
S F_m \le A_c \le S F_M 
\end{eqnarray}
will be identifiable. Smaller particles will be detected if their intrinsic scattering coefficient $c$ is sufficiently large, or by matching the beam surface as well as possible to their diameter. The choice of laser (nano- or micro-) can best match the range of feedback levels as discussed in section~\ref{simp}.

Fig.~\ref{fbck}b shows the radius $r$ of the detectable particle as a function of its scattering coefficient, at constant feedback fraction $F$.  As expected, we see that the detectable particle size decreases as the scattering coefficient increases, for a given amount of feedback.

\subsection{Detection considerations}\label{dets}
Up until here, we have not considered the connection between the particle's speed and the detection scheme.  The finite beam size $S$ introduces a constraint onto the number of points which can be acquired during transit and on the amount of information needed to reconstruct the autocorrelation function.  As we are considering an implementation of principle, rather than an actual setup, we are going to define the boundaries which match realistic constraints, without specific reference to a particular implementation.  We set the minimum number $N_m$ of measurements, on a single particle, to determine the autocorrelation with sufficient precision.  The detection bandwidth, $\nu_B$, determines the data acquisition speed, thus the measurement interval $T_m = \frac{N_m}{\nu_B}$. For a measurement path $L_m$, the maximum particle speed will be determined by

\begin{eqnarray}
v_M & = & \frac{L_m}{T_m} \, , \\
& = & \frac{\nu_B L_m}{N_m} \, .
\end{eqnarray}

In order to fix the ideas, we may assume $\nu_B = 10^8 Hz$, $L_m = 100 \mu m$ and $N_m = 10^4$, thus obtaining maximum speed $v_M \approx 1 m \, s^{-1}$. This speed is already sufficiently large for a gas dynamic setting and can be increased, for a microlaser, by up to two orders of magnitude (cf. later). In microfluidic measurements -- a very broad potential field of sensing application -- the order of magnitude of the particle speed is up to $10^{-4} m \, s^{-1}$, several orders of magnitude below the estimated limit. Thus, we can consider the constraint on speed as being quite easily fulfilled by the setup.

Fig.~\ref{fbck}c shows the values of the maximum speed as a function of $L_m$, for two different sets of parameters: measurement number $N_m$ and bandwidth $\nu_B$. As clearly visible, it is possible to span a large interval of maximum speed values, thus offering a range of measurement conditions which can satisfy a broad set of requirements.


Considering a circular geometry for the beam shape (Fig.~\ref{setup}) and assuming the measurement path to be the beam's diameter, we can link the maximum speed to the geometrical parameters
\begin{eqnarray}
v_M & = & \frac{\nu_B}{T_m} \sqrt{\frac{S}{\pi}} \, .
\end{eqnarray}

Going back to our example, in microfluidics this does not amounts a limitation, thus the choice of surface size $S$ can be adapted to best match the requirements of the measurement (particle size, Fig.~\ref{fbck}a). For other (e.g., gas flow) configurations, instead beam size and particle speed may have to be chosen compatibly with experimental constraints, if one wants to either increase $N_m$ or increase flow rates.

The determination of the bandwidth, for the previous estimates, is based on the following considerations. We recall the two schemes: linear detection with digitization of the signal and computation of $g^{(2)}(\tau)$ from eq.~(5) or photon counting with the direct acquisition of the autocorrelation from a Time-to-Amplitude-Converter (TAC). The latter scheme (Det. 2) uses two avalanche photodetectors, preceded by a beamsplitter on the common path and by an attenuator (not shown) in each of the separate paths (to ensure the measurement of one photon at a time), connected to a discriminator and TAC: a standard setup~\cite{Morton1968}. Since the photon bursts typically contain $10^2$ photons, the attenuator for the photon counting scheme has to considerably reduce the photon flux, thereby rendering the system insensitive to the background spontaneous emission which arrives between bursts.

For the linear detector bandwidths $\nu_B \approx 10^{10} Hz$ are possible (cf. Section IVA), but require larger photon fluxes, such as those provided by the microlaser. The photon counting technique, instead, works at the level of a single photon (thus the attenuator) but has an effective bandwidth reduced by the recovery time. In such a case the maximum $\nu_B \approx 10^8 Hz$, the (more conservative) value used for our first numerical estimate of the maximum particle speed.

Reliance on the below-threshold spontaneously emitted pulses may suggest a slow and unpredictable sensor response.  While randomness is intrinsic, the timescales on which the photon bursts are emitted~\cite{Wang2020JMO} are short enough not to represent a problem.  Measurements on $\beta \approx10^{-4}$ VCSELs have shown typical repetition times in the nanosecond range. Even though the waiting time is not deterministic, it is reasonable to assume that a $10 ns$ delay may be a worst-case scenario. Compared to the residence time of a scatterer in the beam, estimated at the beginning of this section, this does not represent a problem. Smaller lasers (higher $\beta$) may give very large bursts which occur less frequently when the feedback is very small or inexistent (cf. Fig.~\ref{TDynamics}e,f).  However, even in this case the frequency is not a major issue, since smaller peaks (up to 100 photons per pulse) occur with a period of at least $10 ns$ for the lowest feedback (Fig.~\ref{TDynamics}f) and rapidly grow.  Comparison to the typical bandwidths in photon counting (typical dead time of the order of $10 ns$), confirms the viability of the scheme.

\subsection{Particle density and statistics}\label{manyp}
If the particles are monodisperse -- thus the amount of feedback per particle, $F_p$, is well-controlled --, then the system can measure the simultaneous presence of several particles in the beam.  The estimated $F$, obtained from $g^{(2)}$, is nothing else than $F = n F_p$, where $n$ is the number of particles contributing to feedback.  The case of a polydisperse sample is more complex to treat, but information can be collected even in this case if the particles' size statistics is known.  For a probability distribution in size, denoted by $\mathcal{P}(A)$, one can statistically assume that each particle size contributes with its corresponding $F_p(A)$ for the fraction of presence in the sample.

In any case, the amount of feedback $F$ is a measure of the backscattering that reaches the laser, thus of the {\it effective surface} of the sample which has ``reflected back'' part of the laser output.  If the flow is stationary (i.e., constant amount of particles flowing in and out of the measurement area) there is no longer a limitation on the particle speed, since entering particles replace, in average, the exiting ones. There is also no limitation to the beam size $S$ since the transit time no longer enters into the reasoning. Finally, the beam size $S$ ceases to be an adjustment parameter to determine the feedback coefficient $F$: if the sample is homogeneous, any modification of $S$ is equivalent to modifying the number of particles (assumed sufficiently large to avoid feeling their discreteness).  Thus, control in $F$ can only be regained by changing the concentration (now far from the single-particle regime) or by attenuating the amount of feedback, if the latter is too strong.  This can be easily done by optically ensuring that only a fraction of the fed-back photons is coupled into the cavity.

\section{Conclusion}
In conclusion, we have theoretically investigated the photon statistics and nonlinear dynamics of a nanolaser ($\beta = 10^{-1}$) and of a microlaser ($\beta = 10^{-4}$) at the lasing threshold, under the influence of feedback. In this special region, we use the photon bursts emitted and reinjected into the laser to estimate the amount of feedback through the second-order autocorrelation function. Based on the high sensitivity of the autocorrelation function with feedback, we propose two kinds of sensing methods. Using micro- and nanolasers with the two sensing schemes enables the coverage of a broad range of feedback levels, with very good expected accuracy. The principle for an experimental implementation is discussed together with its characteristics. By properly tuning the speed of particles, we found the small laser based sensors can effectively detect the flowing particles. These compact and flexible sensors hold great potential in broad applications, for instance as cell-counting devices coupled to other functions such as in~\cite{Mejia2021}. General particle counting applications in fluids (gases or liquids) can be envisaged.

\section*{Acknowledgment}
T. W. acknowledges the financial support from the National Natural Science Foundation of China (Grant No. 61804036), Zhejiang Province Commonweal Project (Grant No. LGJ20A040001), G. W. acknowledges National Key R $\&$ D Program Grant (Grant No. 2018YFE0120000), and Zhejiang Provincial Key Research $\&$ Development Project Grant (Grant No. 2019C04003). The Authors are grateful to two Referees for insightful comments and detailed questions which have led to a considerable improvement of manuscript quality.


\begin{thebibliography}{00}

\bibitem{Brown1956} R. Hanbury-Brown and R. Q. Twiss, ``Correlation between photons in two coherent beams of light,'' \textit{Nature} vol. 177, pp. 27-29, 1956.

\bibitem{Hanbury-Brown1956} R. Hanbury-Brown and R. Q. Twiss, ``A test of a new type of stellar interferometer on sirius,'' \textit{Nature}, vol. 178, pp. 1046-1048, 1956.

\bibitem{Glauber1963L} R. J. Glauber, ``Photon correlations'', \textit{Phys. Rev. Lett.}, vol. 10, pp. 84-87, 1963.

\bibitem{Glauber1963} R. J. Glauber, ``The quantum theory of optical coherence,'' \textit{Phys. Rev.}, vol. 130, pp. 2529-2539, 1963.

\bibitem{Mandel1965} L. Mandel and E. Wolf, ``Coherence properties of optical fields,'' \textit{Rev. Mod. Phys.}, vol. 37, pp. 231-287, 1965.

\bibitem{Paul1982} H. Paul, ``Photon antibunching,'' \textit{Rev. Mod. Phys.}, vol. 54, pp. 1061-1102, 1982.

\bibitem{Leuchs1986} G. Leuchs, ``Photon statistics, antibunching and squeezed states,'' pp. 329-360 in Frontiers of Nonequilibrium Statistical Physics, G.T. Moore and M.O. Scully, eds. NATO ASI Series B, vol. 135 (Plenum Press, New York, 1986).

\bibitem{Shapiro2012} J. H. Shapiro and R. W. Boyd, ``The physics of ghost imaging,'' \textit{Quantum Inf. Process.}, vol. 11, pp. 949-993, 2012.

\bibitem{Tao2021} M. Tao, X. Gong, J. Guan, J. Song, Z. Song, X. Li, S. Guo, J. Chen, S.Yu, and F. Gao, ``Ghost imaging by a proportional parameter to filter bucket data,'' \textit{Appl. Sci.}, vol. 11, 227, 2021.

\bibitem{Lawson2020} L.M. Lawson, ``Coherent states of position-dependent mass in strong quantum gravitational background fields,'' arXiv:2012.10551. 2020.

\bibitem{Schulz-Dubois1987} E.O. Schulz-Dubois, ``Sizing of microscopic particles by photon correlation,'' pp. 110-129 in {\it Optical Metrology}, ed. O.D.D. Soares, NATO ASI Series E: vol 131. (Martinus Nijhoff Publishers, Dordrecht, NL, 1987).

\bibitem{Schaetzel1991} K. Sch\"atzel and E.0. Schulz-DuBois ``Improvements of photon correlation techniques,'' \textit{Infrared Phys.} vol. 32, 409, 1991.

\bibitem{Schulz-Dubois1983} E. O. Schulz-DuBois, ed., Photon correlation techniques in fluid mechanics,
Sprinqer Verlag, 1983.

\bibitem{Degiorgio1977} V. Degiorgio, ``Photon correlation techniques,'' pp. 142-163 in {\it Photon Correlation Spectroscopy and Velocimetry}, ed. H.Z. Cummings et al. (Springer Science and Business Media, New York, 1977).

\bibitem{Mirhosseini2020} M. Mirhosseini, A. Sipahigil, M. Kalaee, and O. Painter, ``Superconducting qubit to optical photon transduction,'' \textit{Nature}, vol. 588, pp. 599-603, 2020.

\bibitem{Kurochin2020} V. L. Kurochkin, A. V. Khmelev, I. V. Petrov, A. V. Miller, A. A. Feimov, V. F. Mayboroda, M. Y. Balanov, V. V. Krushinsky, A. A. Popov, and Y. V. Kurochkin, ``Registration of the quantum state of a single photon to create a satellite quantum network,'' \textit{J. Phys.: Conf. Series}, vol. 1680, 012031, 2020.

\bibitem{Silberhorn2007} Ch. Silberhorn, ``Detecting quantum light,'' \textit{Cont. Phys.}, vol. 48, pp. 143-156, 2007.

\bibitem{Morais2020} L. A. Morais, T. Weinhold, M. P. de Almeida, A. Lita, Th. Gerrits, S. W. Nam, A. G. White, and G. Gillett, ``Precisely determining photon-number in real time,'' arXiv:2012.10158, 2020.

\bibitem{Wiersig2009}
J. Wiersig, C. Gies, F. Jahnke, M. A\ss mann, T. Berstermann, M. Bayer, C. Kistner, S. Reitzenstein, C. Schneider, S.H\"ofling, A. Forchel, C. Kruse, J. Kalden, and D. Hommel, ``Direct observation of correlations between individual photon emission events of a microcavity laser,'' \textit{Nature}, vol. 460, pp. 245-249, 2009.

\bibitem{Hill2014} M. T. Hill and M. C. Gather, ``Advances in small lasers,'' \textit{Nature Phot.}, vol. 8, pp. 908-918, 2014.

\bibitem{Ma2019} R. M. Ma and R. F. Oulton, ``Applications of nanolasers,'' \textit{Nature Nanotech.}, vol. 14, pp. 12-22, 2019.

\bibitem{Smit2012} M. Smit, J.J.G.M. van der Tol, and M. Hill, ``Moore's law in photonics,'' \textit{Laser \& Photonics Rev.}, vol. 6, pp. 1-13, 2012.

\bibitem{Miller2017} D.A.B. Miller, ``Attojoule optoelectronics for low-energy information processing and communications,'' \textit{J. Lightwave Technol.}, vol. 35, pp. 346-396, 2017.

\bibitem{Lebreton2013L} A. Lebreton, I. Abram, R. Braive, I. Sagnes, I. Robert- Philip, and A. Beveratos, ``Unequivocal differentiation of coherent and chaotic light through interferometric photon correlation measurements,'' \textit{Phys. Rev. Lett.}, vol. 110, 163603, 2013.

\bibitem{Lebreton2013} A. Lebreton, I. Abram, R. Braive, I. Sagnes, I. Robert-Philip, and A. Beveratos, ``Theory of interferometric photon-correlation measurements: differentiating coherent from chaotic light,'' \textit{Phys. Rev. A}, vol. 88, 013801, 2013.

\bibitem{Pan2016} S.H. Pan, Q. Gu, A. El Amili, F. Vallini, and Y. Fainman, ``Dynamic hysteresis in a coherent high-$\beta$ nanolaser,'' \textit{Optica}, vol. 3, pp. 1260-1265, 2016.

\bibitem{Ota2017} Y. Ota, M. Kakuda, K. Watanabe, S. Iwamoto, and Y. Arakawa, ``Thresholdless quantum dot nanolaser,'' \textit{Opt. Express}, vol. 25, pp. 19981-19994, 2017.

\bibitem{Wang2015} T. Wang, G.P. Puccioni, and G.L. Lippi, ``Dynamical buildup of lasing in mesoscale devices,'' \textit{Sci. Rep.}, vol. 5, 15858, 2015.

\bibitem{Wang2020a} T. Wang, D. Aktas, O. Alibart , \'E. Picholle, G. P. Puccioni, S. Tanzilli, and G. L. Lippi, ``Superthermal-light emission and nontrivial photon statistics in small lasers,'' \textit{Phys. Rev. A}, vol. 101, 063835, 2020.

\bibitem{Wang2020JMO} T. Wang, G.P. Puccioni, and G.L. Lippi, ``Photon bursts at lasing onset and modelling issues in micro-VCSELs'', \textit{J. Mod. Optics}, vol. 67, pp. 55-68, 2020.

\bibitem{Lippi2021} G. L. Lippi, ``Amplified spontaneous emission in micro- and nanolasers,'' \textit{Atoms}, vol. 9, 6 (2021).

\bibitem{Redlich2016} Ch. Redlich, B. Lingnau, S. Holzinger, E. Schlottmann, S. Kreinberg, Ch. Schneider, M. Kamp, S. H\"ofling, J. Wolters, S. Reitzenstein, and K. L\"udge, ``Mode-switching induced super-thermal bunching in quantum-dotmicrolasers'', \textit{New J. Phys.}, vol. 18, 063011, 2016.

\bibitem{Marconi2018} M. Marconi, J. Javaloyes, Ph. Hamel, F. Raineri, A. Levenson, and A.M. Yacomotti, ``Far-from-equilibrium route to superthermal light in bimodal nanolasers,'' \textit{Phys. Rev. X}, vol. 8, 011013, 2018.

\bibitem{Panajotov2012} K. Panajotov, M. Sciamanna, M.A. Arteaga, and H. Thienpont, ``Optical
feedback in Vertical-Cavity Surface-Emitting lasers,'' \textit{IEEE J.Sel. Topics Quantum Electron.}, vol. 19, 1700312, 2012.

\bibitem{Wang2019} T. Wang, X. Wang, Z. Deng, J. Sun, G.P. Puccioni, G. Wang, and G.L. Lippi, ``Dynamics of a micro-VCSEL operated in the threshold region under low-level optical feedback,'' \textit{J. Sel. Topics Quantum Electron.}, vol. 25, 1700308, 2019.

\bibitem{Wang2020}
T. Wang, Z. L. Deng, J. C. Sun, X. H. Wang, G. P. Puccioni, G. F. Wang, and G. L. Lippi, ``Photon statistics and dynamics of nanolasers subject to intensity feedback,'' \textit{Phys. Rev. A}, vol. 101, 023803, 2020.

\bibitem{Tredicce1985} J.R. Tredicce, F.T. Arecchi, G.L. Lippi, and G.P. Puccioni, ``Instabilities in Lasers with an Injected Signal'', \textit{J. Opt. Soc. Am. B}, vol. 2, 173-183, 1985.

\bibitem{Roy2009} K. Roy-Choudhury, S. Haas, and A.F.J. Levi, ``Quantum fluctuations in small lasers'', \textit{Phys. Rev. Lett.}, vol. 102, 053902, 2009.

\bibitem{Roy2010} K. Roy-Choudhury, and A.F.J. Levi, ``Quantum fluctuations in very small laser diodes'', \textit{Phys. Rev. A}, vol. 81, 013827, 2010.

\bibitem{Vallet2019} A. Vallet, L. Chusseau, F. Philippe, and A. Jean-Marie, ``Low-dimensional Systems and Nanostructures Markov model of quantum fluctuations at the transition to lasing of semiconductor nanolasers'', \textit{Physica E}, vol. 105, pp. 97-104, 2019.

\bibitem{Takemura2019} N. Takemura, M. Takiguchi, and and M. Notomi, ``Probing the Ginzburg-Landau potential for lasers using higher-order photon correlations'', arXiv:1908.08679.

\bibitem{Takemura2021} N. Takemura, M. Takiguchi, and M. Notomi, ``Low-and high-$\beta$ lasers in the class-A limit: photon statistics, linewidth, and the laser-phase transition analogy'', \textit{J. Opt. Soc. Am. B}, vol. 38, pp. 699-710, 2021.

\bibitem{Puccioni2015}
G. P. Puccioni and G. L. Lippi, ``Stochastic simulator for modeling the transition to lasing,'' \textit{Optics Express}, vol. 23, pp. 2369-2374, 2015.

\bibitem{Druten2000} N. J. van Druten, Y. Lien, C. Serrat, S. S. R. Oemrawsingh, M. P. van Exter, and J. P. Woerdman, ``Laser with thresholdless intensity fluctuations,'' \textit{Phys. Rev. A}, vol. 62, 053808, 2000.

\bibitem{Hofmann2000} H.F. Hofmann and O. Hess, ``Coexistence of thermal noise and squeezing in the intensity fluctuations of small laser diodes,'' \textit{J. Opt. Soc. Am. B}, vol. 17, pp. 1926-1933, 2000.

\bibitem{Bjork1991} G. Bj\"ork, Y, Yamamoto, S. Machida, and K. Igeta, ``Modification of spontaneous emission rate in planar dielectric microcavity structures,'' \textit{Phys. Rev. A}, vol. 44, 669, 1991.

\bibitem{Elvira2011} D. Elvira, X. Hachair, V. B. Verma, R. Braive, G. Beaudoin, I. Robert-Philip, I. Sagnes, B. Baek, S. W. Nam, E. A. Dauler, I. Abram, M. J. Stevens, and A. Beveratos, ``Higher-order photon correlations in pulsed photonic crystal nanolasers,'' \textit{Phys. Rev. A}, vol. 84, 061802(R), 2011.

\bibitem{Foster1998}
G. T. Foster, S. L. Mielke, and L. A. Orozco, ``Intensity correlations of a noise-driven diode laser,'' \textit{J. Opt. Soc. Am. B}, vol. 15, pp. 2646-2653, 1998.

\bibitem{Jin1994}
R. Jin, D. Boggavarapu, M. Sargent III, P. Meystre, H. M. Gibbs, and G. Khitrova, ``Photon-number correlations near the threshold of microcavity lasers in the weak-coupling regime,'' \textit{Phys. Rew. A}, vol. 49, pp. 4038-4042, 1994.

\bibitem{Ates2007}
S. Ates, S.M. Ulrich, P. Michler, S. Reitzenstein, A. L\"offler, and A. Forchel, ``Coherence properties of high-$\beta$ elliptical semiconductor micropillar lasers,'' \textit{Appl. Phys. Lett.}, vol. 90, 161111, 2007.

\bibitem{Rice1994}
P. R. Rice and H. J. Carmichael, ``Photon statistics of a cavity-QED laser: A comment on the laser–phase-transition analogy,'' \textit{Phys. Rev. A}, vol. 50, 4318, 1994.

\bibitem{Jahnke2016} F. Jahnke, Ch. Gies, M. A\ss mann, M. Bayer, H.A.M. Leymann, A. Foerster, J. Wiersig, Ch. Schneider, M. Kamp, and S. H\"ofling, ``Giant photon bunching, superradiant pulse emission and excitation trapping in quantum-dot nanolasers'', Nature Commun. vol. 7, 11540, 2016.


\bibitem{Siegman1986} A. E. Siegman, \textit{Lasers} (University Science Books, Mill Valley, CA, USA, 1986).

\bibitem{Morton1968} G. A. Morton, \textit{Appl. Opt.}, vol. 7, pp. 1-10, 1968.

\bibitem{Mejia2021} J. Mej\'{\i}a Morales, B. Hammarstr\"om, G.L. Lippi, M. Vassalli, and P. Glynne-Jones, ``Acoustofluidic phase microscopy in a tilted segmentation-free configuration'', Biomicrofluidics vol. 15, 014102, 2021.



\end{thebibliography}
\end{document}